\documentclass[12pt]{article}
\usepackage{amsfonts}
\usepackage{amsmath}

\begin{document}
\title{\bf
Integrability of generalized (matrix) Ernst equations in string theory}
\author{
G.~A.~Alekseev${}^1$\footnote{e-mail: G.A.Alekseev@mi.ras.ru}
\\
{\normalsize\emph{${}^1$ Steklov Mathematical Institute, Russian Academy of Sciences}}
\\[-0.5ex]
{\normalsize\emph{Gubkina St. 8, 119991, Moscow, Russia }}
}
\date{}
\maketitle
\abstract{The integrability structures of the matrix generalizations of the Ernst equation for Hermitian or complex symmetric $d\times d$-matrix Ernst potentials are elucidated. These equations arise in the string theory as the equations of motion for a truncated bosonic parts of the low-energy effective action respectively for a dilaton and $d\times d$ - matrix of moduli fields or for a string gravity model with a scalar (dilaton) field, $U(1)$ gauge vector field and an antisymmetric 3-form field, all depending on two space-time coordinates only. We construct the corresponding  spectral problems based on the overdetermined $2d\times 2d$-linear systems with a spectral parameter and the universal (i.e. solution independent) structures of the canonical Jordan forms of their matrix coefficients. The additionally imposed conditions of existence for each of these systems of two matrix integrals with appropriate symmetries provide a specific (coset) structures of the related matrix variables. An equivalence of these spectral problems to the original field equations is proved and some approach for construction of multiparametric families of their solutions is envisaged.}

\bigskip
\noindent
\hbox{\it Keywords:} \quad  {\footnotesize Ernst equations, string gravity, integrability, spectral problems, monodromy}

\subsection*{Introduction}
Integrability properties of various known today integrable reductions of Einstein's field equations in General Relativity for the field configurations  admitting two-dimensional groups of space-time symmetries have been studied using different forms of the reduced dynamical equations, each found most convenient in different contexts. The construction for these equations of equivalent spectral problems, prolongation structures, representations of the infinite-dimensional algebra of internal symmetries opened different ways for applications to the analysis of these equations of various powerful ideas and methods of the modern theory of nonlinear integrable systems such as the inverse scattering methods and soliton techniques, group-theoretic approach and B\"acklund transformations, algebro-geometric methods and finite-gap solutions, etc.

Although the integrability of space-time symmetry reduced Einstein equations was discovered first for vacuum gravitational fields using a matrix sigma-model-like form of these equations \cite{Belinski-Zakharov:1978}, \cite{Belinski-Zakharov:1979}, the Ernst equations (or generalized Ernst equations) represent very compact forms of the dynamical parts of various integrable reductions of Einstein's field equations, providing convenient tools for the analysis of their structures and internal symmetries as well as for development of various solution generating methods. Originally, the Ernst equation for stationary axisymmetric vacuum gravitational fields was derived by F.J. Ernst \cite{Ernst:1968a} as a nonlinear equation for a complex scalar potential $\mathcal{E}$. A similar reduction of electrovacuum Einstein - Maxwell equations for stationary axisymmetric fields leads to a system of two quasi-linear equations for two scalar complex Ernst potentials $\mathcal{E}$, $\Phi$ \cite{Ernst:1968b}. This system reduces to the mentioned above vacuum Ernst equation provided the electromagnetic Ernst potential $\Phi$ vanishes. Besides these elliptic equations, the similar reductions of vacuum Einstein equations and electrovacuum  Einstein - Maxwell equations for the fields depending on time and one spatial coordinate also lead to the so called hyperbolic Ernst equations. Farther generalization of these integrable reductions of the field equations of classical General Relativity had been found for Einstein - Maxwell - Weyl equations for gravitational, electromagnetic and massless two-component spinor fields \cite{Alekseev:1983}.

In the recent years much attention has been attracted by various low-energy effective string and string gravity models which bosonic mode dynamics is described by some generalized Einstein equations for a system of gravitational and massless scalar, vector and tensor fields with a very specific coupling between them. In the simplest models with only some of the bosonic modes of the string considered to be excited and in the presence of two commuting space-time isometries the reduced dynamical equations of the corresponding truncated effective field theory have been found equivalent to some matrix generalizations of the vacuum Ernst equation. In particular, for the reduced dynamical equations of a string model with a scalar (dilaton), $d\times d$ - matrix of moduli fields and vanishing all gauge fields the generalized Ernst potential is a Hermitian $d\times d$ - matrix \cite{Kumar-Ray:1995}. In another case, a gravity model with one scalar field (dilaton), one $U(1)$ gauge vector field and an antisymmetric 3-form field whose dual is expressed using a pseudoscalar (axion) field\footnote{This model is known also as the Einstein - Maxwell - dilaton - axion theory or simply EMDA model. However, it seems useful to mention that this name and its abbreviature are as convenient as missleading because, as one can see easily, the field equations of this model for vanishing axion and dilaton fields don't reduce to the standard Einstein - Maxwell equations of General Relativity but to those with essential additional constraints.} is described by generalized (matrix) Ernst equations for a complex symmetric $2\times 2$ matrix Ernst potential \cite{Bakas:1994, Gal'tsov-et-alii}.  The similarity of these generalized matrix equations to the vacuum Ernst equation which is known to be integrable, as well as the large groups of internal symmetries found for these equations had created reasonable sureness for many authors that these matrix equations are also completely integrable. However, it seems that the theory of these equations has not been developed enough to elucidate their rich (viz. integrable) internal structure and to use it for constructions of various nontrivial solutions. The purpose of this paper is to piece out this shortage and to reformulate the mentioned above generalized Ernst equations in terms of equivalent matrix spectral problems and to envisage a general approach for construction of multiparametric families of their solutions.

\subsection*{Matrix Ernst equations}
For the field configurations depending on two of the four space-time coordinates the dynamical equations of the string models mentioned in the Introduction can be reduced to the following complex matrix equations
\begin{equation}\label{ErnstEqs}
\begin{array}{c}
    2{\cal E}_{\xi\eta}-\displaystyle\frac 1{\xi-\eta}({\cal E}_\xi-{\cal E}_\eta)-
{\cal E}_\xi\cdot(\hbox{Re}{\cal E})^{-1}\cdot{\cal E}_\eta - {\cal E}_\eta\cdot(\hbox{Re}{\cal E})^{-1}\cdot{\cal E}_\xi=0\\[3ex]
(a)\quad {\cal E}^\dagger={\cal E}\qquad\hbox{or}\qquad (b)\quad {\cal E}^\ast={\cal E}
\end{array}
\end{equation}
where ${}^\dagger$ denotes the Hermitian conjugation and ${}^\ast$ is a transposition of the matrices, $\mathcal{E}$ is a complex $d\times d$-matrix Ernst potential which is Hermitian (case (a)) or symmetric (case (b)). The case (b) with $d=1$ corresponds to the original vacuum Ernst equation. The dimension $d$ of the matrix $\mathcal{E}$ should be $d=8$ for $2D$ heterotic string model \cite{Kumar-Ray:1995} or $d=2$ for the $4D$ gravity model with axion, dilaton and one $U(1)$ vector fields \cite{Gal'tsov-et-alii}, however below we leave this natural (positive integer) parameter to be unspecified.

The space-time coordinates $\xi$ and $\eta$ in (\ref{ErnstEqs}) are real null-cone coordinates in the hyperbolic case (the Ernst potential $\mathcal{E}$ and  all field variables in this case are functions of time and one spatial coordinate) or they are complex conjugated to each other in the elliptic case (the Ernst potential $\mathcal{E}$ and all field variables in this case depend on two spatial coordinates only). To distinguish these cases we use the parameters $\epsilon$, $j$:
$$\epsilon=\left\{\begin{array}{rl}
1\hskip-1ex &\hskip1ex \hbox{-- the hyperbolic case,}\\
-1\hskip-1ex &\hskip1ex \hbox{-- the elliptic case,}
\end{array}\quad j=\right\{\begin{array}{ll}
1&\hbox{for $\epsilon=1$,}\\
i&\hbox{for $\epsilon=-1$.}
\end{array}
$$

The considerations below concern the space of all local solutions of (\ref{ErnstEqs}) which are assumed to be holomorphic functions of $\xi$ and $\eta$ in some local region near a chosen ``initial'' or ``reference'' point $(\xi_0,\eta_0)$. It is assumed also that using the existing gauge freedom the Ernst potentials at this initial point are reduced to a ``standard'' value:
\begin{equation}\label{NormE}
    \mathcal{E}(\xi_0,\eta_0)=\mathcal{G}_0,\qquad \mathcal{G}_0=\hbox{diag}\{\varepsilon_1,\,\varepsilon_2,\,\ldots,\,
\varepsilon_d\,\},\qquad \varepsilon_k^2=1.
\end{equation}
Later we assume this ``normalization'' for all solutions. The choice of  $\varepsilon_k$ (the signature of
$\mathcal{G}_0$) is the same for all solutions of a given model, but it can be model-dependent and different for hyperbolic and elliptic cases.

\subsection*{Spectral problems for $d\times d$-matrix equations (\ref{ErnstEqs})}
For construction of the spectral problems for the equations (\ref{ErnstEqs}) we use a ``selfdual''  overdetermined linear systems with a spectral parameter which generalize a system found for vacuum Einstein equations (in another context) by Kinnersley \cite{Kinnersley-Chitre:1978}. For various integrable reductions of Einstein's field equations the systems of this type have been used by different authors (see, for example, \cite{Hauser-Ernst:2001}, \cite{Alekseev:1987}, \cite{Sibgatullin:1984} and the references therein). For the case  (b) and $d=2$ such form of the spectral problem was suggested in the paper \cite{Alekseev-Yurova:2004}. Now we consider the most general spectral problem for $2d\times 2d$-matrices
\begin{equation}\label{Matrices}
    \mathbf{\Psi}(\xi,\eta,w),\hskip1ex  \mathbf{U}(\xi,\eta),  \hskip1ex \mathbf{V}(\xi,\eta),\hskip1ex  \mathbf{W}(\xi,\eta,w)
\end{equation}
where $w$ is a spectral parameter. This includes
the following parts.

\paragraph{$\underline{\hbox{\it The overdetermined linear system.}}$} This is the linear system for $\mathbf{\Psi}$ with the condition that its matrix coefficients should possess the universal (i.e. solution independent) canonical (Jordan) forms:
\begin{equation}\label{LinSys}
\begin{array}{l}
{}\\[-6ex]
\left\{\begin{array}{l}
2 i(w-\xi)\partial_\xi{\bf \Psi}={\bf U}{\bf \Psi},\\
2 i(w-\eta)\partial_\eta{\bf \Psi}={\bf V}{\bf \Psi}\\
\end{array}\right\Vert
\left.\begin{array}{l}
{}\\
{\bf U}_0={\cal F}_+^{-1} {\bf U}{\cal F}_+ =\hbox{diag}\,\{i,\ldots,i,0,\ldots,\,0\},\\
{\bf V}_0={\cal F}_-^{-1} {\bf V}{\cal F}_- =\hbox{diag}\, \{\underset d{\underbrace{i,\ldots,i}},\underset d{\underbrace{0,\ldots,\,0}}\}.
\end{array}\right.
\end{array}
\end{equation}
where we do not impose any constraints on the transformation matrices ${\cal F}_\pm$ besides a condition of their existence. We require also the existence for the system (\ref{LinSys}) of two matrix integrals of special structures.

\paragraph{$\underline{\hbox{\it The matrix integrals in the Hermitian case (a).}}$}
In this case one of the integrals, denoted farther as $\mathbf{K}(w)$, should be symmetric and satisfy the conditions
\begin{equation}\label{WIntegrala}
{\bf \Psi}^\ast {\bf W}{\bf \Psi}={\bf K}(w), \hskip1ex {\bf
K}^\ast(w)={\bf K}(w), \quad \displaystyle\frac {\partial {\bf
W}}{\partial w}=4i\underset {(a)}{\bf \Omega},\quad
\underset {(a)}{\bf \Omega}=
\begin{pmatrix} 0&I\\I&0
\end{pmatrix},
\end{equation}
while the other integral, denoted below as $\mathbf{L}(w)$, is Hermitian and possessing the structure
\begin{equation}\label{OmegaIntegrala}
    \Sigma(\xi,\eta,w) {\bf \Psi}^\dagger \underset {(a)}{\bf \Omega}{\bf \Psi}={\bf L}(w),\quad  {\bf L}^\dagger(w)={\bf L}(w),\quad
\Sigma^2\equiv (w-\xi)(w-\eta),
\end{equation}
where $\mathbf{\Psi}^\dagger(\xi,\eta,w)\equiv \overline{\mathbf{\Psi}^\ast(\xi,\eta,\overline{w})}$, the matrices ${\bf K}(w)$ and ${\bf L}(w)$ are coordinate independent functions of $w$, $\Sigma$ is an auxiliary scalar function, ${\bf \Omega}$ is a symmetric constant matrix and $I$ is a unit $d\times d$-matrix.

\paragraph{$\underline{\hbox{\it The matrix integrals in the symmetric case (b).}}$} In this case, the first integral should be Hermitian and determined by the expressions
\begin{equation}\label{WIntegralb}
{\bf \Psi}^\dagger {\bf W}{\bf \Psi}={\bf K}(w), \qquad {\bf
K}^\dagger(w)={\bf K}(w), \qquad \displaystyle\frac {\partial {\bf
W}}{\partial w}=4i\underset {(b)}{\bf \Omega}
\end{equation}
while the second integral should be antisymmetric and it takes  the form
\begin{equation}\label{OmegaIntegralb}
\Sigma(\xi,\eta,w){\bf \Psi}^\ast \underset {(b)}{\bf \Omega}{\bf \Psi}={\bf L}(w),\qquad  {\bf L}^\ast(w)=-{\bf L}(w),\qquad
\underset {(b)}{\bf \Omega}=
\begin{pmatrix} 0&I\\-I&0
\end{pmatrix}
\end{equation}
where the matrix $I$ and the function $\Sigma$ are the same as in (\ref{WIntegrala}) and (\ref{OmegaIntegrala}).

\paragraph{$\underline{\hbox{\it Gauge transformations and normalization of solutions.}}$}
The spectral problems described above,
admit two groups of pure gauge transformations. One of them consists of the transformations of the form $\mathbf{\Psi}\to\mathbf{\Psi} \mathbf{C}(w)$, where $\mathbf{C}(w)$ is an arbitrary non-degenerate matrix depending on the spectral parameter only. This gauge freedom can be used for a normalization of the function $\mathbf{\Psi}$ at the chosen initial point $(\xi_0,\eta_0)$, where we have normalized the Ernst potential by the condition (\ref{NormE}). Choosing  $\mathbf{C}(w)=\mathbf{\Psi}^{-1}(\xi_0,\eta_0,w)$, we achieve without any loss of generality
\begin{equation}\label{NormPsi}
\mathbf{\Psi}(\xi_0,\eta_0,w)=\mathbf{I}
\end{equation}
where $\mathbf{I}$ is a $2d\times 2d$ unit matrix. This specifies the values of the integrals\footnote{For simplicity, here and below in most expressions we do not supply the matrix $\mathbf{\Omega}$ and various functions by the case corresponding suffices (a) or (b).}:
\begin{equation}\label{IntNorms}
\mathbf{K}(w)=\mathbf{W}_0(w),\qquad \mathbf{L}(w)=\Sigma(\xi_0,\eta_0,w) \mathbf{\Omega},
\end{equation}
where $\mathbf{W}_0(w)$ is the value of $\mathbf{W}(\xi,\eta,w)$ at the initial point. The transformations of the second group don't violate the condition (\ref{NormPsi}). They are determined by the following actions
\begin{equation}\label{Atransform}
\mathbf{\Psi}\to\mathbf{A}\,\mathbf{\Psi}\,\mathbf{A}^{-1},
\quad\begin{array}{lcl}
\mathbf{U}\to\mathbf{A}\,\mathbf{U}\,\mathbf{A}^{-1}&& \mathbf{W}\to\mathbf{A}^\dagger{}^{-1}\,\mathbf{W}\, \mathbf{A}^{-1}
\\
\mathbf{V}\to\mathbf{A}\,\mathbf{V}\,\mathbf{A}^{-1}&& \mathbf{W}_0\to\mathbf{A}^\dagger{}^{-1}\,\mathbf{W}_0\, \mathbf{A}^{-1}
\end{array}
\end{equation}
where a constant $2d\times 2d$-matrix $\mathbf{A}$ should satisfy the invariance conditions $\mathbf{A}^\dagger\mathbf{\Omega} \mathbf{A}=\mathbf{\Omega}$ and $\mathbf{A}^\ast\mathbf{\Omega} \mathbf{A}=\mathbf{\Omega}$. These imply that the matrix $\mathbf{A}$ should be real and possess for the cases (a) and (b) respectively the structure
\begin{equation}\label{Astruc}
\underset{(a)}{\mathbf{A}}=\begin{pmatrix}
 (L^\ast)^{-1}+\omega_2 L\,\omega_1& \omega_2 L\\[1ex]
L\,\omega_1 & L
\end{pmatrix},\quad
\underset{(b)}{\mathbf{A}}=\begin{pmatrix}
(L^\ast)^{-1}+\delta_2 L\,\delta_1& \delta_2 L\\[1ex]
L\, \delta_1 & L
\end{pmatrix}
\end{equation}
where $L$ is an arbitrary non-degenerate real $d\times d$ matrix, $\omega_1$, $\omega_2$ are real antisymmetric $d\times d$-matrices and $\delta_1$, $\delta_2$ are real symmetric $d\times d$-matrices.
In accordance with (\ref{IntNorms}) the normalization (\ref{NormPsi}) corresponds to a special forms of the matrix integrals
\begin{equation}\label{NormIntegrals}
    \begin{array}{lclclcl} (a):&&\mathbf{\Psi}^\ast\,\mathbf{W}\,\mathbf{\Psi}=\mathbf{W}_0(w),
&&(b):&&\mathbf{\Psi}^\dagger\, \mathbf{W}\,\mathbf{\Psi}=\mathbf{W}_0(w),
  \\[2ex]
&&\mathbf{\Psi}^\dagger\, \mathbf{\Omega}\, \mathbf{\Psi}=\lambda_+^{-1}\lambda_-^{-1} \, \mathbf{\Omega}&&
&&\mathbf{\Psi}^\ast\, \mathbf{\Omega}\,\mathbf{\Psi}=\lambda_+^{-1}\lambda_- ^{-1}\mathbf{\Omega}
    \end{array}
\end{equation}
where the scalar functions $\lambda_+(\xi,w)=\sqrt{(w-\xi)/(w-\xi_0)}$, $\lambda_-(\eta,w)=\sqrt{(w-\eta)/(w-\eta_0)}$ and $\lambda_+(\xi,w=\infty)=1$, $\lambda_-(\eta,w=\infty)=1$.
As it follows from (\ref{WIntegrala}) and
(\ref{WIntegralb}), the matrix $\mathbf{W}(\xi,\eta,w)$ and therefore, its initial value $\mathbf{W}_0(w)\equiv \mathbf{W}(\xi_0,\eta_0,w)$
are linear functions of $w$ with the case-dependent symmetry:
\begin{equation}\label{Wstruct}
\begin{array}{l}
\mathbf{W}=4i (w-\beta)\mathbf{\Omega}+\mathbf{G}\\[0ex]
\mathbf{W}_0=4i (w-\beta_0)\mathbf{\Omega}+\mathbf{G}_0
\end{array}
\hskip1ex
\begin{array}{llll}
(a)\,:&\mathbf{G}=\mathbf{G}^\ast &(b)\,:&\mathbf{G}=\mathbf{G}^\dagger\\[0ex]
&\mathbf{G}_0=\mathbf{G}_0^\ast & &\mathbf{G}_0=\mathbf{G}_0^\dagger
\end{array}
\end{equation}
where the terms with  $\beta\equiv (\xi+\eta)/2$ and $\beta_0\equiv (\xi_0+\eta_0)/2$ were introduced for later convenience.
The gauge transformations (\ref{Atransform}), (\ref{Astruc})  can be used for reducing of the matrices $\mathbf{G}_0$ to some standard values, say
\begin{equation}\label{G0struc}
 (a)\,:\quad \mathbf{G}_0=4\begin{pmatrix} -\epsilon\alpha_0^2\mathcal{G}_0,&0\\[1ex]
0&\mathcal{G}_0\end{pmatrix},\qquad
(b)\,:\quad \mathbf{G}_0=-4\begin{pmatrix} \epsilon\alpha_0^2\mathcal{G}_0,&0\\[1ex]
0&\mathcal{G}_0\end{pmatrix}
\end{equation}
where $\alpha_0=(\xi_0-\eta_0)/2j$ and the $d\times d$-matrix $\mathcal{G}_0$ is chosen to coinciding with the initial value for the Ernst potential  defined in (\ref{NormE}).

\subsection*{The spectral problems versus the Ernst equations.}

$\underline{\hbox{\textsc{Theorem}}}$ {\it The spectral problem (\ref{Matrices}), (\ref{LinSys}), (\ref{WIntegrala}), (\ref{OmegaIntegrala}) is equivalent to the  Ernst equation (\ref{ErnstEqs}) with a Hermitian Ernst potential and the spectral problem (\ref{Matrices}), (\ref{LinSys}), (\ref{WIntegralb}), (\ref{OmegaIntegralb}) is equivalent to the  Ernst equation (\ref{ErnstEqs}) with a complex symmetric Ernst potential matrices.}
\medskip

\paragraph{$\underline{\hbox{\it The condition of existence of the integral $\mathbf{L}(w)$.}}$} Differentiating (\ref{OmegaIntegrala}), (\ref{OmegaIntegralb}) and using (\ref{DifConjugations}) and (\ref{LinSys}) we obtain that the existence of the integral $\mathbf{L}(w)$ leads to the relations\footnote{Differentiating the expressions (\ref{OmegaIntegrala}) and (\ref{OmegaIntegralb}) with respect to $\xi$ and $\eta$ we have to take into account that $\xi$ and $\eta$ are real in the hyperbolic case and
they are complex conjugated to each other in the elliptic case. Therefore, for $\epsilon=1$ and $\epsilon=-1$ we obtain respectively
\begin{equation}\label{DifConjugations}
\left\{\begin{array}{l}
\partial_\xi({\bf \Psi}^\dagger)=-{\bf \Psi}^\dagger {\bf U}^\dagger/2 i(w-\xi)\\
\partial_\eta({\bf \Psi}^\dagger)=-{\bf \Psi}^\dagger {\bf V}^\dagger/2 i(w-\eta)
\end{array}\right.\qquad
\left\{\begin{array}{l}
\partial_\xi({\bf \Psi}^\dagger)=-{\bf \Psi}^\dagger {\bf V}^\dagger/2 i(w-\xi)\\
\partial_\eta({\bf \Psi}^\dagger)=-{\bf \Psi}^\dagger {\bf U}^\dagger/2 i(w-\eta)
\end{array}\right.
\end{equation}
}:
\begin{equation}\label{UVsymmetry}
\begin{array}{l}
(a)\hskip2ex\begin{array}{l}
\epsilon=1\,:\\[2ex]
{}
\end{array}\hskip0ex
\left\{\begin{array}{l}
\mathbf{\Omega}\mathbf{U}-\mathbf{U}^\dagger \mathbf{\Omega}=i \mathbf{\Omega}\\[1ex]
\mathbf{\Omega}\mathbf{V}-\mathbf{V}^\dagger \mathbf{\Omega}=i \mathbf{\Omega}
\end{array}\right.\hskip0ex
\begin{array}{l}
\epsilon=-1\,:\\[2ex]
{}
\end{array}\hskip0ex
\left\{\begin{array}{l}
\mathbf{\Omega}\mathbf{U}-\mathbf{V}^\dagger \mathbf{\Omega}=i \mathbf{\Omega}\\[1ex]
\mathbf{\Omega}\mathbf{V}-\mathbf{U}^\dagger \mathbf{\Omega}=i \mathbf{\Omega}
\end{array}\right.\\[5ex]
(b)\hskip3ex\epsilon=\pm 1\,:\qquad
\mathbf{\Omega}\mathbf{U}+\mathbf{U}^\ast \mathbf{\Omega}=i \mathbf{\Omega},\qquad
\mathbf{\Omega}\mathbf{V}+\mathbf{V}^\ast \mathbf{\Omega}=i \mathbf{\Omega}
\end{array}
\end{equation}

\paragraph{$\underline{\hbox{\it The condition of existence of the integral $\mathbf{K}(w)$.}}$}   Differentiating (\ref{WIntegrala})\\ and (\ref{WIntegralb})  and using (\ref{DifConjugations}), (\ref{LinSys}), we obtain for the case (a):
\begin{equation}\label{UVGrelationsa}
\begin{array}{ll}
\partial_\xi\mathbf{G}= 2 i\mathbf{\Omega}-2\mathbf{\Omega} \mathbf{U}-2\mathbf{U}^\ast \mathbf{\Omega},&
(\mathbf{G}+4i j\alpha \mathbf{\Omega})\mathbf{U}=-\mathbf{U}^\ast (\mathbf{G}+4i j\alpha \mathbf{\Omega}),\\[1ex]
\partial_\eta\mathbf{G}= 2 i\mathbf{\Omega}-2\mathbf{\Omega} \mathbf{V}-2\mathbf{V}^\ast \mathbf{\Omega},&
(\mathbf{G}-4i j\alpha \mathbf{\Omega})\mathbf{V}=-\mathbf{V}^\ast (\mathbf{G}-4i j\alpha \mathbf{\Omega}).
\end{array}
\end{equation}
where $2 j\alpha=\xi-\eta$. For the case (b) with $\epsilon=1$ we have the similar relations
\begin{equation}\label{UVGrelationsb1}
\begin{array}{ll}
\partial_\xi\mathbf{G}= 2 i\mathbf{\Omega}-2\mathbf{\Omega} \mathbf{U}+2\mathbf{U}^\dagger \mathbf{\Omega}&
(\mathbf{G}+4i\alpha \mathbf{\Omega})\mathbf{U}=\mathbf{U}^\dagger (\mathbf{G}+4i\alpha \mathbf{\Omega})\\[1ex]
\partial_\eta\mathbf{G}= 2 i\mathbf{\Omega}-2\mathbf{\Omega} \mathbf{V}+2\mathbf{V}^\dagger \mathbf{\Omega}&
(\mathbf{G}-4i\alpha \mathbf{\Omega})\mathbf{V}=\mathbf{V}^\dagger (\mathbf{G}-4i\alpha \mathbf{\Omega})
\end{array}
\end{equation}
and for the case (b) with $\epsilon=-1$ we have
\begin{equation}\label{UVGrelationsbm1}
\begin{array}{ll}
\partial_\xi\mathbf{G}= 2 i\mathbf{\Omega}-2\mathbf{\Omega} \mathbf{U}+2\mathbf{V}^\dagger \mathbf{\Omega}&
(\mathbf{G}-4\alpha \mathbf{\Omega})\mathbf{U}=\mathbf{V}^\dagger (\mathbf{G}-4\alpha \mathbf{\Omega})\\[1ex]
\partial_\eta\mathbf{G}= 2 i\mathbf{\Omega}-2\mathbf{\Omega} \mathbf{V}+2\mathbf{U}^\dagger \mathbf{\Omega}&
(\mathbf{G}+4\alpha \mathbf{\Omega})\mathbf{V}=\mathbf{U}^\dagger (\mathbf{G}+4\alpha \mathbf{\Omega})
\end{array}
\end{equation}

\paragraph{$\underline{\hbox{\it  Structure of $\mathbf{G}$, $\mathbf{U}$, $\mathbf{V}$ implied by existence of matrix integrals.}}$}\hskip-1ex If we express now the products $\mathbf{\Omega}\mathbf{U}$ and $\mathbf{\Omega}\mathbf{V}$ from (\ref{UVsymmetry}), substitute them into the corresponding equations (\ref{UVGrelationsa}) or (\ref{UVGrelationsb1}), (\ref{UVGrelationsbm1}) with derivatives of $\mathbf{G}$ and apply a transposition to the equations obtained, taking into account the symmetries (\ref{Wstruct}) of $\mathbf{G}$, we obtain for both cases (a) and (b) the relations
\begin{equation}\label{UVG}
\begin{array}{llll}
\epsilon=1\,:&
\partial_\xi\mathbf{G}=-2\mathbf{\Omega}(\mathbf{U}+ \overline{\mathbf{U}}) &\epsilon=-1\,:&
\partial_\xi\mathbf{G}=-2\mathbf{\Omega}(\mathbf{U}+ \overline{\mathbf{V}})\\[1ex]
&
\partial_\eta\mathbf{G}=-2\mathbf{\Omega}(\mathbf{V}+ \overline{\mathbf{V}}) & &
\partial_\eta\mathbf{G}=-2\mathbf{\Omega}(\mathbf{V}+ \overline{\mathbf{U}})
\end{array}
\end{equation}
From these relations for $\epsilon=1$ or using a complex conjugation and taking into account the rules (\ref{DifConjugations}) for $\epsilon=-1$, we arrive at
$\partial_\xi(\mathbf{G}-\overline{\mathbf{G}})=0$ and $\partial_\eta(\mathbf{G}-\overline{\mathbf{G}})=0$. In view of  (\ref{Wstruct}) and (\ref{G0struc}) these last conditions mean the reality of $\mathbf{G}$.
Thus, in both cases (a) and (b) we obtain
\begin{equation}\label{GReality}
\mathbf{G}=\mathbf{G}^\ast,\qquad \mathbf{G}=\overline{\mathbf{G}}.
\end{equation}

We note now that for the case (a) the equations in the left column of (\ref{UVGrelationsa}) and in the case (b) the equations in the left column in (\ref{UVGrelationsb1}) or (\ref{UVGrelationsbm1}) lead to the following expressions for the matrices $\mathbf{U}$ and $\mathbf{V}$
\begin{equation}\label{UVab}
\begin{array}{lll}
(a)\,:&\mathbf{U}=-\dfrac14 \mathbf{\Omega}\,\partial_\xi\mathbf{G}+ \dfrac i2\mathbf{I}+i\mathbf{\Omega}\,X_+,&
\mathbf{V}=-\dfrac14 \mathbf{\Omega}\,\partial_\eta\mathbf{G}+ \dfrac i2\mathbf{I}+i\mathbf{\Omega}\,X_-\\[1ex]
(b)\,:&\mathbf{U}=\dfrac14 \mathbf{\Omega}\,\partial_\xi\mathbf{G}+ \dfrac i2\mathbf{I}+i\mathbf{\Omega}\,Y_+,&
\mathbf{V}=\dfrac14 \mathbf{\Omega}\,\partial_\eta\mathbf{G}+ \dfrac i2\mathbf{I}+i\mathbf{\Omega}\,Y_-
\end{array}
\end{equation}
where $X_\pm$ are arbitrary antisymmetric and $Y_\pm$ are arbitrary symmetric $2d\times 2d$-matrices, and $X_\pm$, as well as $Y_\pm$, are real in the case $\epsilon=1$ or complex conjugated to each other in the case $\epsilon=-1$.

Now we have to substitute these expressions into the corresponding algebraic equations of the right hand side columns in (\ref{UVGrelationsa}) and (\ref{UVGrelationsb1}), (\ref{UVGrelationsbm1}). Then separating the real and imaginary parts of these equations for the case (a) or taking certain linear combinations of these equations and their complex conjugations for the case (b) we obtain  $\partial_\xi(\mathbf{G}\mathbf{\Omega}{\mathbf{G}}
\pm 16\epsilon\alpha^2\mathbf{\Omega}) =0$ and $\partial_\eta(\mathbf{G}\mathbf{\Omega}{\mathbf{G}}
\pm 16\epsilon\alpha^2\mathbf{\Omega}) =0$, where the upper signs correspond to the case (a) and the lower ones -- to the case (b). Using the expressions (\ref{Wstruct}) and (\ref{G0struc}) we obtain from the last conditions the relations
\begin{equation}\label{GOmegaG}
(a):\quad\mathbf{G}\mathbf{\Omega}{\mathbf{G}}
=-16\epsilon\alpha^2\mathbf{\Omega},\qquad\qquad
(b):\quad\mathbf{G}\mathbf{\Omega}{\mathbf{G}}
=16\epsilon\alpha^2\mathbf{\Omega}.
\end{equation}
These conditions mean that the matrices $\mathbf{G}$ should possess the structures
\begin{equation}\label{GStruc}
\underset{(a)}{\mathbf{G}}=4\begin{pmatrix}
-\mathcal{A} \mathcal{G} \mathcal{A}-\epsilon\alpha^2 \mathcal{G}^{-1}&  -\mathcal{A}\mathcal{G} \\[1ex]
\mathcal{G}\,\mathcal{A}&\mathcal{G}
\end{pmatrix},\hskip1ex
\underset{(b)}{\mathbf{G}}=-4\begin{pmatrix}
\mathcal{S} \mathcal{G} \mathcal{S}+\epsilon\alpha^2 \mathcal{G}^{-1}&  \mathcal{S}\mathcal{G} \\[1ex]
\mathcal{G}\,\mathcal{S}&\mathcal{G}
\end{pmatrix}
\end{equation}
where $\mathcal{G}$, $\mathcal{S}$ are symmetric and $\mathcal{A}$ is an antisymmetric real $d\times d$-matrices. From the equations (\ref{UVGrelationsa}), (\ref{UVGrelationsb1}), (\ref{UVGrelationsbm1}) we obtain also
\begin{equation}\label{omdel}
\begin{array}{ll}
(a)\,:&\mathbf{G}-2 j\alpha \partial_\xi\mathbf{G}+ \mathbf{G} \mathbf{\Omega}\,X_+-X_+ \mathbf{\Omega} \mathbf{G}=0,\\
&\mathbf{G}-2 j\alpha \partial_\eta\mathbf{G}+ \mathbf{G} \mathbf{\Omega}\,X_--X_- \mathbf{\Omega} \mathbf{G}=0\\[1ex]
(b)\,:&\mathbf{G}-2 j\alpha \partial_\xi\mathbf{G}+ \mathbf{G}\mathbf{\Omega}\,Y_+-Y_+ \mathbf{\Omega} \mathbf{G}=0,\\
&\mathbf{G}-2 j\alpha \partial_\eta\mathbf{G}+ \mathbf{G} \mathbf{\Omega}\,Y_--Y_- \mathbf{\Omega} \mathbf{G}=0,
\end{array}
\end{equation}
The equations (\ref{omdel}) can be simplified considerably if we express the matrices $X_\pm$, $Y_\pm$ in the form $X_+= (\partial_\xi\mathbf{G}\mathbf{\Omega}\mathbf{G}- \mathbf{G}\mathbf{\Omega}\partial_\xi\mathbf{G})/{32j\alpha}
+\widetilde{X}_+$, $X_-=(\mathbf{G}\mathbf{\Omega}\partial_\eta\mathbf{G}-
\partial_\eta\mathbf{G}\mathbf{\Omega}\mathbf{G})/{32j\alpha}+ \widetilde{X}_-$, $Y_+=(\partial_\xi\mathbf{G}\mathbf{\Omega} \mathbf{G}-\mathbf{G}\mathbf{\Omega}\partial_\xi\mathbf{G}
)/{32j\alpha}+\widetilde{Y}_+$ and $Y_-= (\mathbf{G}\mathbf{\Omega}\partial_\eta\mathbf{G}-
\partial_\eta\mathbf{G}\mathbf{\Omega}\mathbf{G})/{32j\alpha} +\widetilde{Y}_-$, where the new matrix variables $\widetilde{X}_\pm$ are antisymmetric and $\widetilde{Y}_\pm$ are symmetric $2d\times 2d$-matrices and  all of these matrices are real in the case $\epsilon=1$, but in the case $\epsilon=-1$, $\widetilde{X}_\pm$ as well as $\widetilde{Y}_\pm$ are complex conjugated to each other. If we introduce then the matrices $\mathbf{h}$ defined as
\begin{equation}\label{hdef}
    (a)\,:\quad\mathbf{h}=-\dfrac 14\mathbf{\Omega}\mathbf{G}, \qquad\qquad\qquad (b)\,:\quad\mathbf{h}=\dfrac 14\mathbf{\Omega}\mathbf{G},
\end{equation}
the matrices $\mathbf{U}$ and $\mathbf{V}$ can be expressed in the form
\begin{equation}\label{UVh}
\begin{array}{ll}
\underset{(a)}{\mathbf{U}}=\partial_\xi\mathbf{h}-\dfrac i{j\alpha}\mathbf{h}\,\partial_\xi\mathbf{h}+i\mathbf{\Omega} \,\widetilde{X}_+,
&
\underset{(b)}{\mathbf{U}}=\partial_\xi\mathbf{h}-\dfrac i{j\alpha}\mathbf{h}\,\partial_\xi\mathbf{h}+i\mathbf{\Omega} \,\widetilde{Y}_+,\\[1ex]
\underset{(a)}{\mathbf{V}}=\partial_\eta\mathbf{h}+\dfrac i{j\alpha}\mathbf{h}\,\partial_\eta\mathbf{h}+i\mathbf{\Omega} \,\widetilde{X}_-,&
\underset{(b)}{\mathbf{V}}=\partial_\eta\mathbf{h}+\dfrac i{j\alpha}\mathbf{h}\,\partial_\eta\mathbf{h}+i\mathbf{\Omega} \,\widetilde{Y}_-,
\end{array}
\end{equation}
where the matrices $\widetilde{X}_\pm$ and $\widetilde{Y}_\pm$ should satisfy the relations
\begin{equation}\label{XtYteqs}
\begin{array}{llclcl}
(a)\,:&\mathbf{G}\mathbf{\Omega}\,\widetilde{X}_\pm- \widetilde{X}_\pm\mathbf{\Omega}\,\mathbf{G}=0
&&(b)\,:&
\mathbf{G}\mathbf{\Omega}\,\widetilde{Y}_\pm- \widetilde{Y}_\pm\mathbf{\Omega}\,\mathbf{G}=0
\end{array}
\end{equation}
A general solution of these equations can be expressed in terms of the $d\times d$-matrices introduced in (\ref{GStruc}):
\begin{equation}\label{XtYtsol}
\begin{array}{l}
\widetilde{X}_\pm=\begin{pmatrix}\mathcal{G}^{-1}&-\mathcal{A}\\
0&I\end{pmatrix}
\begin{pmatrix}
-\epsilon\alpha^2 \omega_\pm^\prime
&\omega_\pm^{\prime\prime}\\
\omega_\pm^{\prime\prime}&\omega_\pm^\prime
\end{pmatrix}
\begin{pmatrix}\mathcal{G}^{-1}&0\\
\mathcal{A}&I\end{pmatrix}\\[3ex]
\widetilde{Y}_\pm=\begin{pmatrix}\mathcal{G}^{-1}&\mathcal{S}\\
0&I\end{pmatrix}
\begin{pmatrix}
\epsilon\alpha^2 \delta_\pm
&\omega_\pm\\
-\omega_\pm&\delta_\pm^\prime
\end{pmatrix}
\begin{pmatrix}\mathcal{G}^{-1}&0\\
\mathcal{S}&I\end{pmatrix}
\end{array}
\end{equation}
where $\omega_+^\prime$, $\omega_-^\prime$,  as well as $\omega_+^{\prime\prime}$, $\omega_-^{\prime\prime}$
and $\omega_+$, $\omega_-$ are the pairs of antisymmetric and $\delta_+$, $\delta_-$ is a pair of symmetric arbitrary real (for $\epsilon=1$) or complex conjugated to each other (for $\epsilon=-1$) $d\times d$-matrices. Thus, the expressions (\ref{GStruc}), (\ref{UVh}) and (\ref{XtYtsol}) for both cases (a) and (b) represent the general structures of the matrices $\mathbf{G}$, $\mathbf{U}$ and $\mathbf{V}$ which are necessary and, as one can show this easily, sufficient for the existence for the linear system (\ref{LinSys}) of the matrix integrals $\mathbf{K}(w)$ and $\mathbf{L}(w)$.

\paragraph{$\underline{\hbox{\it The conditions for canonical forms of the matrices $\mathbf{U}$ and $\mathbf{V}$.}}$} The constructed above solution  for $\mathbf{U}$ and $\mathbf{V}$ in the case (a) takes the form
\begin{equation}\label{UVexplicita}
\begin{array}{l}
\mathbf{U}=i \mathbf{L}_+
\begin{pmatrix} I+(\omega_+^{\prime\prime}-i j \alpha\omega_+^\prime)\mathcal{G}^{-1}
&\hskip-2ex i(\mathcal{E}_+-i\omega_+^\prime)\\
0&\hskip-2ex\mathcal{G}^{-1}(\omega_+^{\prime\prime}+i j \alpha\omega_+^\prime) \end{pmatrix}\mathbf{L}_+^{-1}\\[3ex]
\mathbf{V}= i \mathbf{L}_-
\begin{pmatrix} I+(\omega_-^{\prime\prime}+i j \alpha\omega_-^\prime)\mathcal{G}^{-1}
&\hskip-2ex i(\mathcal{E}_--i\omega_-^\prime)\\
0&\hskip-2ex\mathcal{G}^{-1}(\omega_-^{\prime\prime}-i j \alpha\omega_-^\prime) \end{pmatrix} \mathbf{L}_-^{-1}\\[3ex]
\hbox{where}\hskip2ex\mathbf{L}_\pm=\begin{pmatrix} I&0\\ -\mathcal{A}\mp i j\alpha \mathcal{G}^{-1}
&I \end{pmatrix},\quad
\begin{array}{l}
\mathcal{E}_+=\mathcal{G}_\xi-\dfrac i{j\alpha}\mathcal{G} \mathcal{A}_\xi\mathcal{G},\\
\mathcal{E}_-=\mathcal{G}_\eta+\dfrac i{j\alpha}\mathcal{G} \mathcal{A}_\eta \mathcal{G}
\end{array}
\end{array}
\end{equation}
and for the case (b) this solution takes a similar form :
\begin{equation}\label{UVexplicitb}
\begin{array}{l}
\mathbf{U}=i \mathbf{L}_+
\begin{pmatrix} I-(\omega_++i j \alpha\delta_+)\mathcal{G}^{-1}
\hskip-3ex & i(\mathcal{E}_+-i\delta_+)\\
0\hskip-3ex & -\mathcal{G}^{-1}(\omega_+-i j \alpha\delta_+) \end{pmatrix} \mathbf{L}_+^{-1}
\\[3ex]
\mathbf{V}=i \mathbf{L}_-
\begin{pmatrix} I-(\omega_--i j \alpha\delta_-)\mathcal{G}^{-1}
\hskip-3ex& i(\mathcal{E}_--i\delta_-)\\
0\hskip-3ex& -\mathcal{G}^{-1}(\omega_-+i j \alpha\delta_-) \end{pmatrix} \mathbf{L}_-^{-1}\\[2ex]
\hbox{where}\hskip2ex\mathbf{L}_\pm=\begin{pmatrix} I&0\\ -\mathcal{S}\mp i j\alpha \mathcal{G}^{-1}
&I \end{pmatrix},\quad
\begin{array}{l}
\mathcal{E}_+=\mathcal{G}_\xi-\dfrac i{j\alpha}\mathcal{G}
\mathcal{S}_\xi \mathcal{G},\\
\mathcal{E}_-=\mathcal{G}_\eta+\dfrac i{j\alpha}\mathcal{G}S_\eta \mathcal{G},
\end{array}
\end{array}
\end{equation}
the suffices $\xi$ and $\eta$ denote the derivatives,
and the properties of $d\times d$-matrices in the above expressions were defined in the previous subsection. Consider now the constraints imposed on the above solution by the condition of a specific structure (\ref{LinSys}) of its canonical Jordan forms. For this we consider the characteristic matrix polynomial equation  which should be satisfied by any matrix with such canonical form:
$$\mathbf{U} \mathbf{U}-i \mathbf{U}=0,\qquad \mathbf{V} \mathbf{V}-i \mathbf{V}=0$$
Substituting here the expressions (\ref{UVexplicita}) and (\ref{UVexplicitb}) presented in the forms
$$\begin{array}{l}
\mathbf{U}=i \mathbf{L}_+\begin{pmatrix} I+\mathcal{X}_+\mathcal{G}^{-1}&\mathcal{Y}_+\\0& \mathcal{G}^{-1}\mathcal{Z}_+\end{pmatrix} \mathbf{L}_+^{-1},\\[2ex]
\mathbf{V}=i \mathbf{L}_-\begin{pmatrix} I+\mathcal{X}_-\mathcal{G}^{-1}&\mathcal{Y}_-\\0& \mathcal{G}^{-1}\mathcal{Z}_-\end{pmatrix} \mathbf{L}_-^{-1}
\end{array}$$
it is easy to find that the equations obtained in this way don't include the  matrices $\mathbf{L}_\pm$ and these equations take the forms
\begin{equation}\label{XYZeqs}
\begin{array}{lccl}
\mathcal{X}_+\mathcal{G}^{-1}\mathcal{X}_+=-\mathcal{X}_+&&& \mathcal{X}_-\mathcal{G}^{-1}\mathcal{X}_-=-\mathcal{X}_-\\
\mathcal{X}_+\mathcal{G}^{-1}\mathcal{Y}_++ \mathcal{Y}_+\mathcal{G}^{-1}\mathcal{Z}_+=0&&& \mathcal{X}_-\mathcal{G}^{-1}\mathcal{Y}_-+ \mathcal{Y}_-\mathcal{G}^{-1}\mathcal{Z}_-=0\\
\mathcal{Z}_+\mathcal{G}^{-1}\mathcal{Z}_+=\mathcal{Z}_+&&& \mathcal{Z}_-\mathcal{G}^{-1}\mathcal{Z}_-=\mathcal{Z}_-
\end{array}
\end{equation}
If we recall now that for the case (a) $\mathcal{X}_\pm$ and $\mathcal{Z}_\pm$ are antisymmetric, and for the case (b) $\mathcal{X}_\pm+\mathcal{Z}_\pm$ and $\mathcal{X}_\pm-\mathcal{Z}_\pm$ are antisymmetric and symmetric $d\times d$-matrices respectively, the separation of symmetric and antisymmetric parts of the equations (\ref{XYZeqs}) leads to the conclusion that $\mathcal{X}_\pm=\mathcal{Z}_\pm=0$. Therefore, the necessary conditions for the matrices $\mathbf{U}$ and $\mathbf{V}$ of the forms (\ref{UVexplicita}) and (\ref{UVexplicitb}) to possess the canonical forms (\ref{LinSys}) are
$$\omega_\pm^\prime=0,\quad \omega_\pm^{\prime\prime}=0,\quad
\omega_\pm=0,\quad
\delta_\pm=0.
$$
It is easy to see that these conditions are also sufficient, and with these conditions the matrices (\ref{UVexplicita}) and (\ref{UVexplicitb})
can be factorized as follows:
\begin{equation}\label{UVF}
\mathbf{U}=\mathcal{F}_+\begin{pmatrix} i I&0\\ 0&0\end{pmatrix} \mathcal{F}_+^{-1}, \hskip1ex
\mathbf{V}=\mathcal{F}_-\begin{pmatrix} i I&0\\ 0&0\end{pmatrix} \mathcal{F}_-^{-1},\hskip1ex
\mathcal{F}_\pm=\mathbf{L}_\pm\begin{pmatrix} I&-i\mathcal{E}_\pm\\ 0&I\end{pmatrix}.
\end{equation}

\paragraph{$\underline{\hbox{\it The Ernst potentials and the Ernst equations.}}$} The integrability conditions of the system (\ref{LinSys})  consist of two matrix equations
\begin{equation}\label{NullCurv}{\bf U}_\eta+{\bf
V}_\xi+\displaystyle\frac 1{2 i j\alpha} [{\bf U},{\bf V}] =
0,\qquad {\bf U}_\eta-{\bf V}_\xi=0,
\end{equation}
where $\alpha\equiv (\xi-\eta)/2 j$. The upper right $d\times d$-block of the last of these equations leads to the condition $\partial_\eta\mathcal{E}_+=\partial_\xi\mathcal{E}_-$ and therefore, there exists the generalized $d\times d$-matrix Ernst potential $\mathcal{E}(\xi,\eta)$ with the properties
$$\partial_\xi\mathcal{E}=\mathcal{E}_+,\quad \partial_\eta\mathcal{E}=\mathcal{E}_-,\qquad \mathcal{R}e\mathcal{E}=\mathcal{G},
\qquad \mathcal{E}(\xi_0,\eta_0)=\mathcal{G}_0.
$$
The expressions (\ref{UVF}) can be written more explicitly in the form
$$
\mathbf{U}=\begin{pmatrix}
i I+\mathcal{E}_+\Omega_+&-\mathcal{E}_+\\
\Omega_+(i I+\mathcal{E}_+\Omega_+)&-\Omega_+\mathcal{E}_+
\end{pmatrix},\hskip1ex
\mathbf{V}=\begin{pmatrix}
i I+\mathcal{E}_-\Omega_-&-\mathcal{E}_-\\
\Omega_-(i I+\mathcal{E}_-\Omega_-)&-\Omega_-\mathcal{E}_-
\end{pmatrix}
$$
where the $d\times d$ matrices $\Omega_\pm$ are defined by the expressions
$$(a)\,:\quad \Omega_\pm =-\mathcal{A}\mp i j\alpha \mathcal{G}^{-1},\qquad (b)\,:\quad \Omega_\pm = -\mathcal{S}\mp i j\alpha \mathcal{G}^{-1}.
$$
The upper right $d\times d$-matrix block
of the first equation in (\ref{NullCurv}) for the matrices $\mathbf{U}$ and $\mathbf{V}$ given just above and with the use of (\ref{UVexplicita}) and (\ref{UVexplicitb}) coincides with the generalized Ernst equations (\ref{ErnstEqs}) with a Hermitian or symmetric $d\times d$-matrix Ernst potential respectively. On the other hand, for any solution of the generalized Ernst equations (\ref{ErnstEqs}) the integrability conditions (\ref{NullCurv}) are satisfied and there exists a solution of the corresponding spectral problem. This completes our proof of equivalence of the spectral problems  constructed above to the corresponding matrix Ernst equations.

\subsection*{Concluding remarks.}
The spectral problems constructed above for the generalized
Ernst equations for Hermitian  or complex symmetric $d\times d$-matrix Ernst potentials provide a general base for generalizations for these equations of various approaches developed earlier for the analysis of the internal structure of other integrable reductions of Einstein`s field equations. The monodromy transform approach \cite{Alekseev:1985}, \cite{Alekseev:1987}, \cite{Alekseev:2000}, \cite{Alekseev:2001} is the most general one, which takes into account many features which are common for all known today integrable reductions of Einstein's field equations. This approach admits a natural generalization to the matrix cases considered here. In particular, it can be shown that any solution of the generalized matrix Ernst equations also can be characterized by a set of functional parameters (depending on the spectral parameter only) which constitute a complete set of the monodromy data on the spectral plane of the corresponding fundamental solution of the linear system (\ref{LinSys}). However, in contrast to the case of the Ernst equation for vacuum gravitational fields, in which these monodromy data can be expressed in terms of a pair of ``projective'' vectors  determined by their affine coordinates in the form $\mathbf{k}_\pm(w)=\{1,\mathbf{u}_\pm(w)\}$, where $\mathbf{u}_\pm(w)$ is a pair of scalar functions of the spectral parameter holomorphic in some local regions of the spectral plane, the monodromy data in the generalized matrix cases are determined by a pair of holomorphic functions which take their values in the complex Grassman space $G_{d,2d}(\mathbb{C})$ and in the affine coordinates they are  determined by two holomorphic $d\times d$-matrix functions $\mathbf{u}_\pm(w)$:
$$\mathbf{k}_\pm(w)=\hskip1ex\begin{array}{ll}
&\hskip1ex\hbox{\Large 0}\\[0ex]
\hbox{\Large 0}&
\end{array}\hskip-12ex
\left.\left(\begin{array}{llll}
\hbox{\small 1}&&&\\
&\ddots&&\\
&&&\hskip-3ex \hbox{\small 1}
\end{array}
\hskip-2ex\right\vert\hskip-1ex
\begin{array}{ccc}
\mathbf{u}_{11}^\pm(w)&\ldots&\mathbf{u}_{1d}^\pm(w)\\
\vdots&\ddots&\vdots\\
\mathbf{u}_{d1}^\pm(w)&\ldots&\mathbf{u}_{dd}^\pm(w)
\end{array}\right)\hskip0ex\begin{array}{l}
(a)\,:\hskip0ex\mathbf{u}_\pm(w)=-\mathbf{u}_\pm^\dagger(w)\\[2ex]
(b)\,:\hskip0ex\mathbf{u}_\pm(w)=\mathbf{u}_\pm^\ast(w)
\end{array}
$$
where the Hermitian conjugation is defined as $\mathbf{u}_\pm^\dagger(w)\equiv \overline{\mathbf{u}_\pm^\ast(\overline{w})}$ in the hyperbolic case ($\epsilon=1$) and $\mathbf{u}_\pm^\dagger(w)\equiv \overline{\mathbf{u}_\mp^\ast(\overline{w})}$ in the elliptic case  ($\epsilon=-1$). The construction of solutions for any given monodromy data can be reduced to solution of some systems of linear singular integral equations of a Cauchy type. These equations  admit infinite hierarchies of explicit solutions as, for example, for rational and analytically matched (viz. $\mathbf{u}_+(w)=\mathbf{u}_-(w)\equiv\mathbf{u}(w)$) monodromy data. However, these questions are expected to be a subject of the subsequent considerations.

\subsection*{Acknowledgments}
The author is grateful to the organizers of the Workshop ``{\it Nonlinear Physics: Theory and Experiment. III}'', Gallipoli, Lecce, Italy, June 24 - July 3, 2004, for invitation and financial support. This work was supported partly by the Russian Foundation for Basic Research (the grants 02-01-00729 and 02-02-17372) and by the programs ``Nonlinear dynamics'' of Russian Academy of Sciences and ``Leading Scientific Schools'' of Russian Federation (the grant NSh-1697.2003.1).


\begin{thebibliography}{99}\itemsep=0pt

\bibitem{Belinski-Zakharov:1978} V. A. Belinski, and  V. E. Zakharov, Sov. Phys. JETP {\bf 48}, 985, (1978).

\bibitem{Belinski-Zakharov:1979} V. A. Belinski, and  V. E. Zakharov, Sov. Phys. JETP {\bf 50}, 1 (1979).

\bibitem{Ernst:1968a} F. J. Ernst, Phys.Rev. {\bf 167} (2), 1175 (1968).

\bibitem{Ernst:1968b} F. J. Ernst, Phys.Rev. {\bf 168} (2), 1415 (1968).

\bibitem{Alekseev:1983} G. A. Alekseev, Sov. Phys. Dokl. (USA) {\bf 28}, 133 (1983).

\bibitem{Kumar-Ray:1995}  A.Kumar and K.Ray, Phys.Lett. {\bf B358} (1995) 223.

\bibitem{Bakas:1994} I.Bakas, Nucl. Phys. {\bf B428}  374 (1994);
Phys.Rev. {\bf D54} (1996) 6424.

\bibitem{Gal'tsov-et-alii}  D. V. Gal'tsov and O. V. Kechkin, Phys. Lett. {\bf B361} 52 (1995); Phys. Rev. {\bf D54} 1656 (1996);  D.V. Gal'tsov and S.A. Sharakin, Phys.Lett. {\bf B399} (1997) 250.

\bibitem{Kinnersley-Chitre:1978} W. Kinnersley and D. M. Chitre, J. Math. Phys. {\bf 19}, 1927 (1978).

\bibitem{Hauser-Ernst:2001} I. Hauser and F. J. Ernst,
Gen. Rel. Grav., {\bf 33}, 195 (2001).

\bibitem{Alekseev:1987} G. A. Alekseev, Proc. Steklov Inst.
Maths. (1988) {\bf 3}, 215.

\bibitem{Sibgatullin:1984} N.R.Sibgatullin, \textit{Oscillations and waves in strong gravitational and electromagnetic fields}, Nauka, Moscow (1984); Engl. transl.: Springer Verlag, Berlin (1991).

\bibitem{Alekseev-Yurova:2004} G. A. Alekseev and M. V. Yurova,
``Integrable structure of the low-energy string gravity equations
in $D=4$ space-times with two commuting isometries '',
Proceedings of International Workshop ``Supersymmetries and Quantum Symmetries'' ((SQS'03), Dubna, Russia, July 24-29, 2003), 159 -- 164 (2004); hep-th/0401077

\bibitem{Alekseev:1985} G. A. Alekseev, Sov. Phys. Dokl. {\bf 30}, 565 (1985).

\bibitem{Alekseev:2000} G.A.Alekseev, ``Monodromy transform approach to solution of some field equations in General Relativity and string heory'', Proceedings of the workshop "Nonlinearity, Integrability and all that: Twenty years after NEEDS'79" (Gallipoli, Lecce, Italy, 1-10 Jul 1999),  p. 12 -- 18,  World Scientific, Singapore  (2000); gr-qc/9911045

\bibitem{Alekseev:2001} G.A.Alekseev, Physica D {\bf 152},  97 (2001);  gr-qc/0001012.



\end{thebibliography}
\end{document}